\documentclass{pazhb_eng}
\usepackage[hyperfootnotes=false]{hyperref}
\usepackage{graphicx}
\usepackage{color}
\usepackage{epsf}
\usepackage[utf8]{inputenc}
\usepackage[T2A]{fontenc}
\usepackage{amsmath}
\usepackage{amsfonts}
\usepackage{amssymb}
\usepackage{epsf}
\usepackage{gensymb}
\usepackage{float}
\usepackage{longtable}
\usepackage{tabularray}
\UseTblrLibrary{booktabs, siunitx}
\SetTblrStyle{contfoot-text}{font=\footnotesize\itshape}
\SetTblrStyle{note}{font=\small}
\usepackage{changes}
\usepackage{wrapfig}
\usepackage{xcolor}
\usepackage{longtable}
\usepackage{xltabular}
\usepackage{changes}
\usepackage{threeparttable}
\usepackage{threeparttablex}
\usepackage{booktabs,supertabular}
\usepackage{multicol}
\usepackage{ltablex}
\usepackage{threeparttable}
\usepackage{threeparttablex}
\usepackage{caption}
\newcommand{\degs}{\ifmmode ^{\circ}\else$^{\circ}$\fi}
\newcommand{\amin}{\ifmmode ^{\prime}\else$^{\prime}$\fi}
\def\asec{\ifmmode ^{\prime\prime}\else$^{\prime\prime}$\fi}

\newcommand{\ergs}{erg~s$^{-1}$}
\newcommand{\flux}{erg~s$^{-1}$~cm$^{-2}$}

\newcommand{\dm}{pc~cm$^{-3}$}

\begin{document}

\journalinfo{2025}{00}{0}{1}[10]
%\UDK{///}

%\maketitle

\title{Identification of Radio and Gamma-ray Pulsars \\ in X-rays Using Data from the SRG/eROSITA \\ All-Sky Survey}
\author{%\bf \hspace{-1.3cm}\copyright\, 2020 г. \ \
Yu.A.~Shibanov\address{1},
A.V.~Karpova\address{1}\email{},
D.A.~Zyuzin\address{1}\email{annakarpova1989@gmail.com (AVK), da.zyuzin@gmail.com (DAZ)},
M.R.~Gilfanov\address{2,3}
\addresstext{1}{Ioffe Institute, Politekhnicheskaya 26, St. Petersburg 194021, Russia}
\addresstext{2}{Space Research Institute, Russian Academy of Sciences, Profsoyuznaya 84/32, 117997 Moscow, Russia}
\addresstext{3}{Max-Planck-Institut f\"ur Astrophysik, Karl-Schwarzschild-Str. 1, D-85741 Garching, Germany}
}

\shortauthor{Shibanov et al.}
\shorttitle{Identification of radio and $\gamma$-ray pulsars}

\begin{abstract} 
Using the data from the all-sky survey in soft X-rays performed by the eROSITA telescope onboard the Spectrum-Roentgen-Gamma observatory we identified known radio and $\gamma$-ray pulsars in the eastern half of the sky.
As a result, new candidate counterparts were found for twelve pulsars of different ages and types at a $\gtrsim 3\sigma$ confidence level. 
A comparable number had been previously identified in the western half of the sky. 
In total, this represents $\approx$12\% of all known pulsars already detected in X-rays. 
For the new counterparts, we provide estimates of their X‑ray fluxes, preliminary characteristics of their X-ray spectra, and brief descriptions of the pulsars' properties. 
In addition, in the eastern half of the sky eROSITA detected 55 pulsars previously identified in X‑rays by other telescopes.

\keywords{pulsars -- sky surveys -- SRG -- eROSITA}
\end{abstract}

\section{Introduction}

Pulsars are highly magnetized, rapidly rotating neutron stars (NSs). 
More than 3800 pulsars\footnote{According to the Australia Telescope National Facility (ATNF) Pulsar catalog \citep{atnf}; \url{https://www.atnf.csiro.au/people/pulsar/psrcat/}.} have been discovered in over half a century of observations. 
Nevertheless, many fundamental questions concerning their internal structure, emission across the whole electromagnetic spectrum, and the link between different NS classes remain unsolved. 
The situation is further complicated by the fact that the vast majority (more than 95\%) of pulsars are observed in the radio, whereas only $\sim$10\% of sources have been detected at higher energies \citep{smith,xu2025}. 
Therefore, the discovery of new objects and the identification of known ones in the optical, X-rays, and $\gamma$-rays are of great importance for studying pulsars' parameters and physics. 
All-sky surveys play a crucial role here, as they often allow one to detect pulsars' candidate counterparts. 
Luminosities in the 0.3--10 keV range can reach $\approx 10^{36}$ \ergs\ (as for the Crab pulsar) while for the faintest pulsar, PSR J0108$-$1431, the luminosity is $\approx 10^{29}$ \ergs\ \citep{xu2025}.

Using X-ray observations, one can independently estimate the distances to pulsars. 
This is important in the case of radio-quiet stars, for which the standard distance estimation using dispersion measure (DM) is impossible. 
Such observations also allow one to study thermal emission from the pulsars' surfaces, which is most often detected in X-rays \citep[e.g.,][]{potekhin2020}. 
Then the mass, radius, and chemical composition of the NS's atmosphere can be estimated. 
When cyclotron lines are detected in the spectrum, the magnetic field strength can be directly inferred \citep[e.g.,][]{gothelf}.

The depth of the new all-sky survey performed by the eROSITA telescope \citep{predehl2020} onboard the Spectrum-Roentgen-Gamma \citep[SRG;][]{sunyaev2021} observatory enables a search for new identifications of known radio and $\gamma$-ray pulsars in soft X-rays. 
This was recently done for the western half of the sky \citep{erosita-grmn}. 
Here we present the results of such a search for the eastern half of the sky, based on positional coincidences and analysis of the X-ray spectra of the detected sources.

%%%%%%%%%%%%%%%%%%%%%%%%%%%%%%%%%%%%%%%%%%%%%%%%%%%%%%%%%%%%%%%%%%%%
%%%%%%%%%%%%%%%%%%%%%%%%%%%%%%%%%%%%%%%%%%%%%%%%%%%%%%%%%%%%%%%%%%%%
%%%%%%%%%%%%%%%%%%%%%%%%%%%%%%%%%%%%%%%%%%%%%%%%%%%%%%%%%%%%%%%%%%%%
%%%%%%%%%%%%%%%%%%%%%%%%%%%%%%%%%%%%%%%%%%%%%%%%%%%%%%%%%%%%%%%%%%%%

\section{The sample of pusars for the X-ray identification}

For identifications, we used the ATNF Pulsar Catalogue (version 2.6.0), which is continuously updated and contains the most comprehensive data on the parameters of more than 3800 objects (2260 in the eastern half of the sky). 
We did not consider pulsars for which the positional uncertainties exceeded 10\asec\ or were not listed in the catalog.
Furthermore, we excluded objects associated with globular clusters, where identifications are challenging due to the insufficient eROSITA angular resolution. 
Additionally, we added pulsars in binary systems recently discovered using the FAST radio telescope \citep{fast}.

%%%%%%%%%%%%%%%%%%%%%%%%%%%%%%%%%%%%%%%%%%%%%%%%%%%%%%%%%%%%%%%%%%%%
%%%%%%%%%%%%%%%%%%%%%%%%%%%%%%%%%%%%%%%%%%%%%%%%%%%%%%%%%%%%%%%%%%%%
%%%%%%%%%%%%%%%%%%%%%%%%%%%%%%%%%%%%%%%%%%%%%%%%%%%%%%%%%%%%%%%%%%%%
%%%%%%%%%%%%%%%%%%%%%%%%%%%%%%%%%%%%%%%%%%%%%%%%%%%%%%%%%%%%%%%%%%%%

\section{X-ray data and the search for candidate counterparts}

%%%%%%%%%%%%%%%%%%%%%%%%%%%%%%%%%%%%%%%%%%%%%%%%%%%%%%%%%%%%%%%%%%%%
%%%%%%%%%%%%%%%%%%%%%%%%%%%%%%%%%%%%%%%%%%%%%%%%%%%%%%%%%%%%%%%%%%%%

\subsection{eROSITA data processing and the search method}

%---------------------------------------------------------------
\begin{figure*} 
\begin{center}
    \includegraphics[width=0.32\textwidth, trim = {0 0.5cm 0 0.3cm},clip]{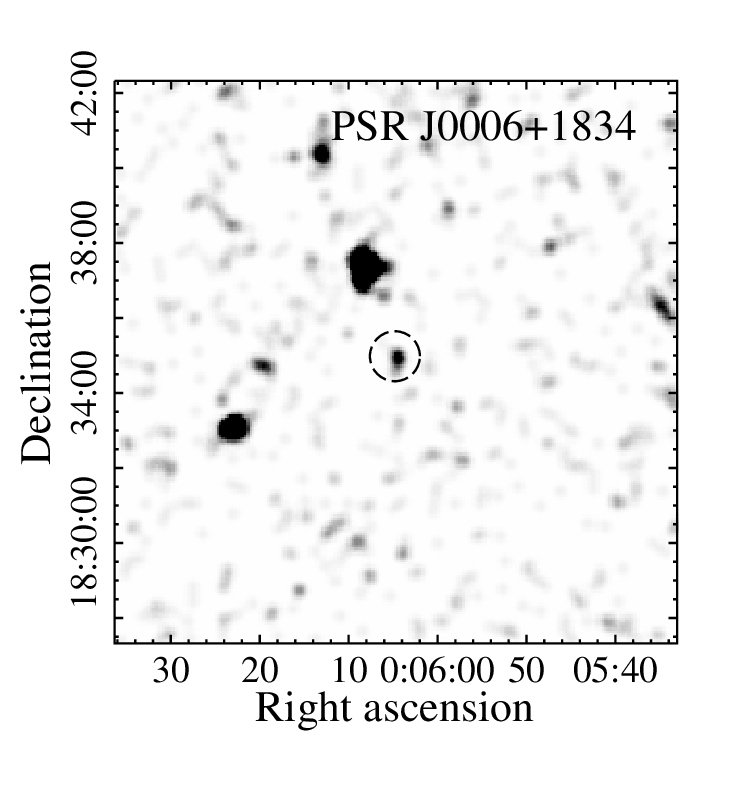}
    \put(-68,26) {\includegraphics[width=0.1\textwidth, clip=]{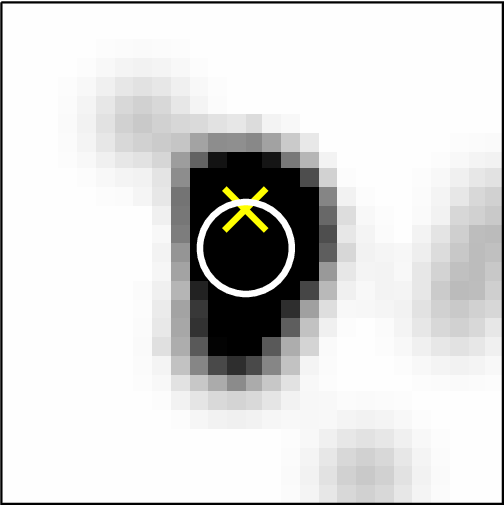}}
    \includegraphics[width=0.32\textwidth, trim = {0 0.5cm 0 0.3cm},clip]{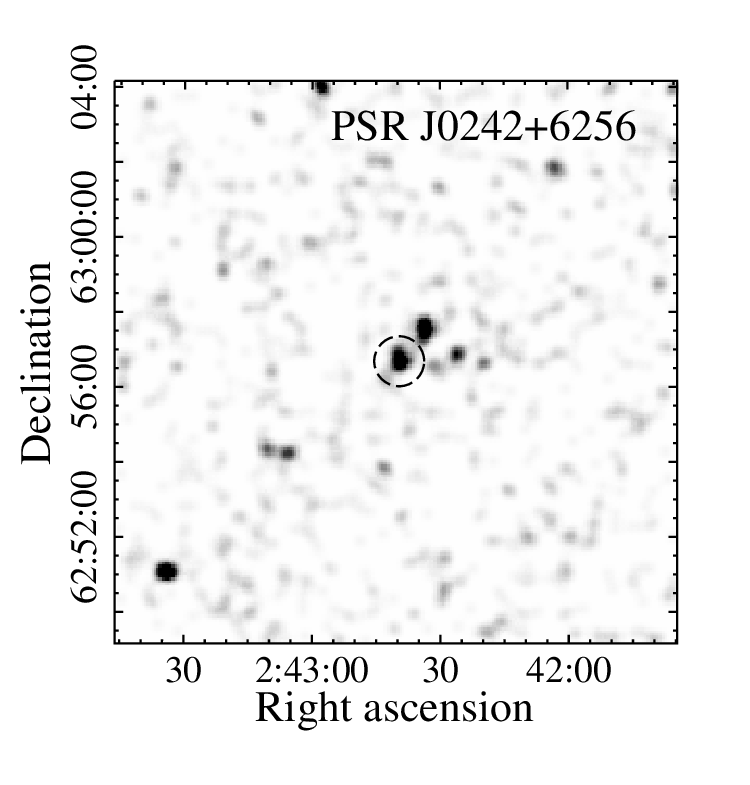}
    \put(-68,26) {\includegraphics[width=0.1\textwidth, clip=]{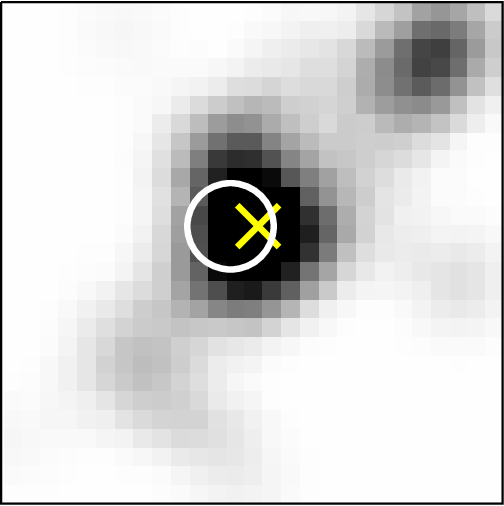}}
    \includegraphics[width=0.32\textwidth, trim = {0 0.5cm 0 0.3cm},clip]{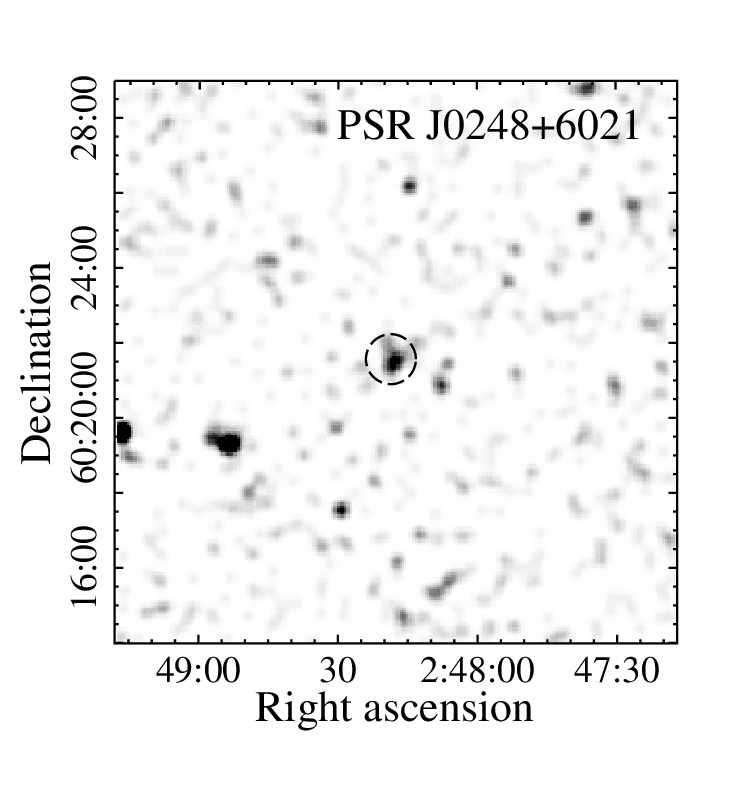}
    \put(-68,26) {\includegraphics[width=0.1\textwidth, clip=]{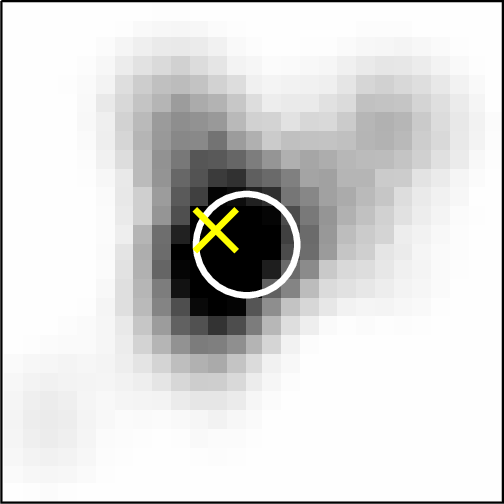}} \\
    \includegraphics[width=0.32\textwidth, trim = {0 0.5cm 0 0.3cm},clip]{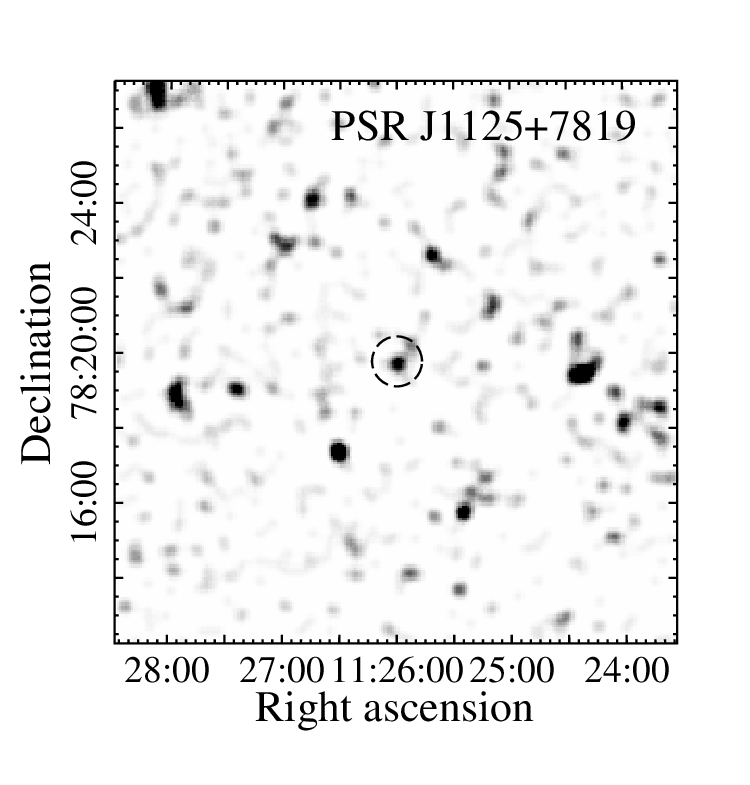}
    \put(-68,26) {\includegraphics[width=0.1\textwidth, clip=]{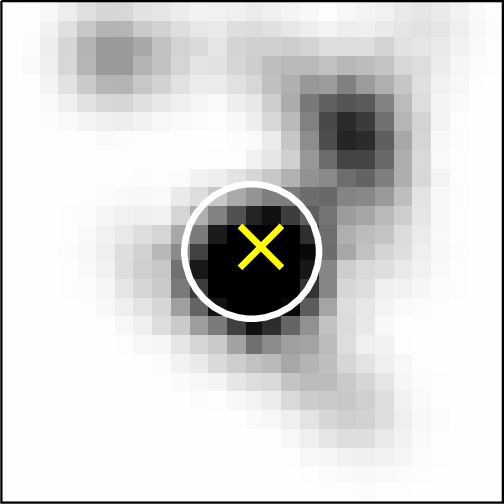}}
    \includegraphics[width=0.32\textwidth, trim = {0 0.5cm 0 0.3cm},clip]{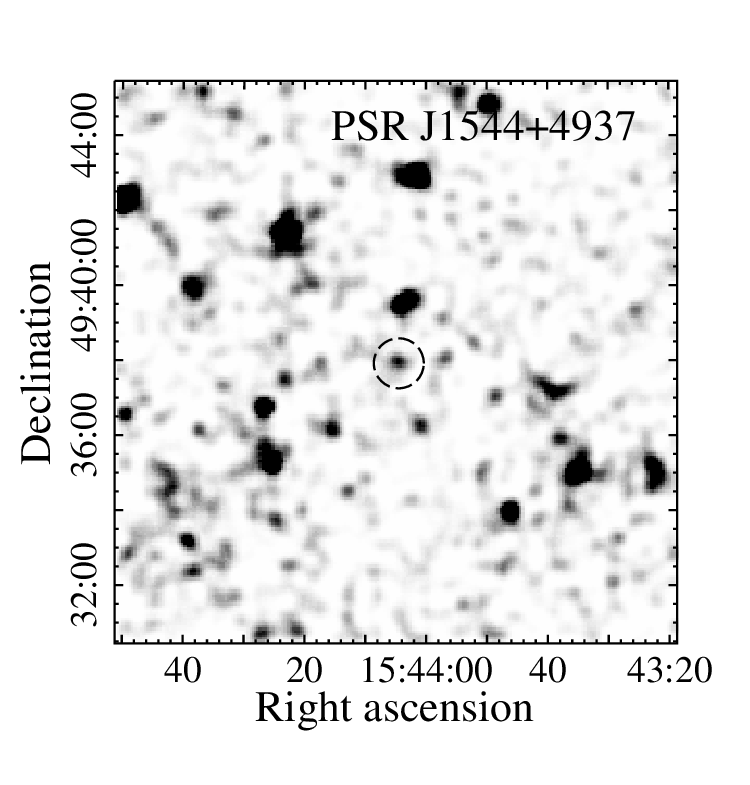}
    \put(-68,26) {\includegraphics[width=0.1\textwidth, clip=]{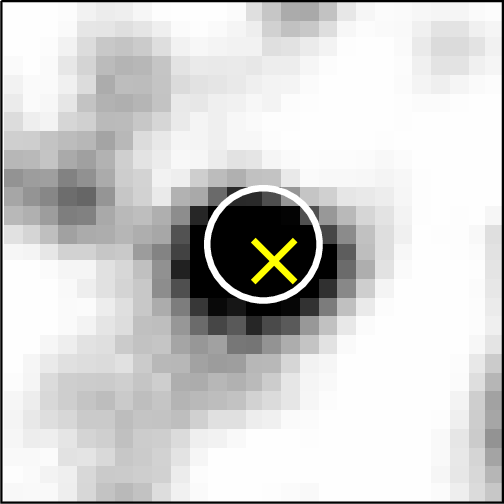}}
    \includegraphics[width=0.32\textwidth, trim = {0 0.5cm 0 0.3cm},clip]{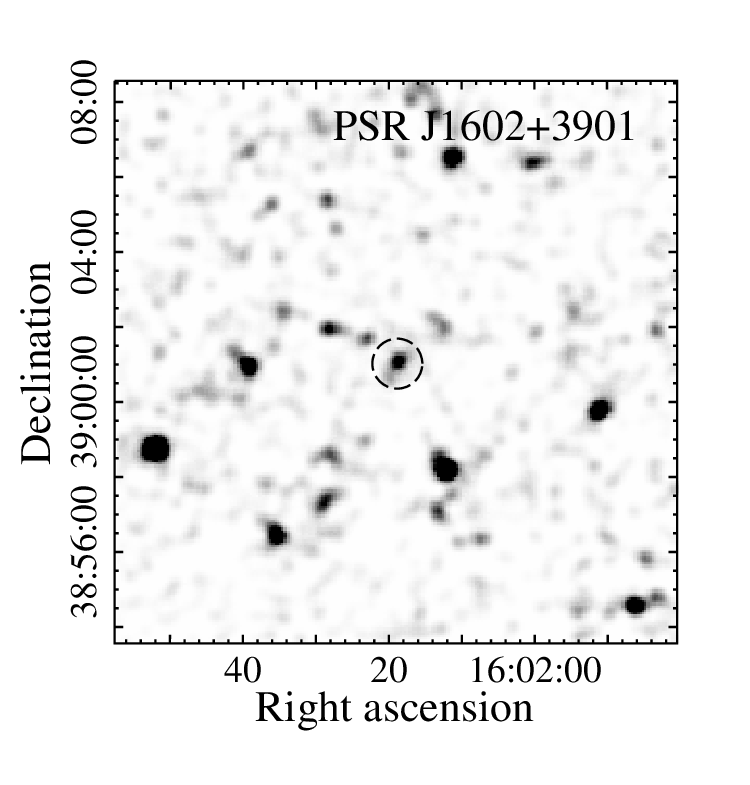}
    \put(-68,26) {\includegraphics[width=0.1\textwidth, clip=]{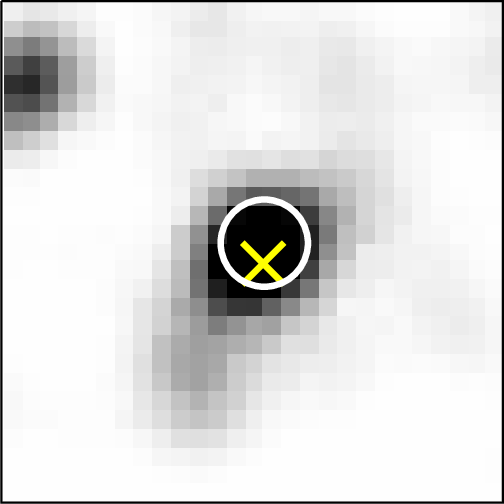}} \\
    \includegraphics[width=0.32\textwidth, trim = {0 0.5cm 0 0.3cm},clip]{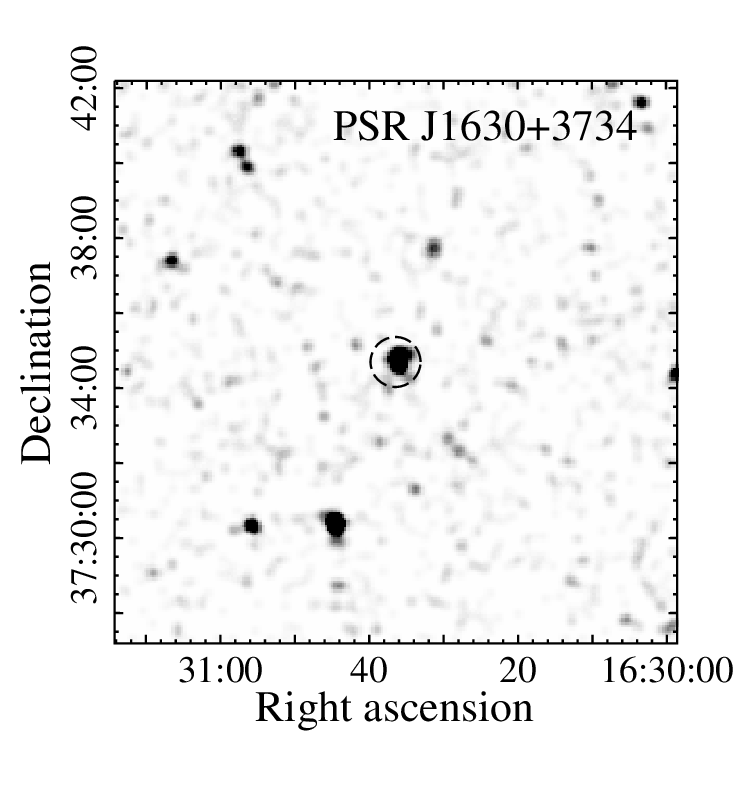}
    \put(-68,26) {\includegraphics[width=0.1\textwidth, clip=]{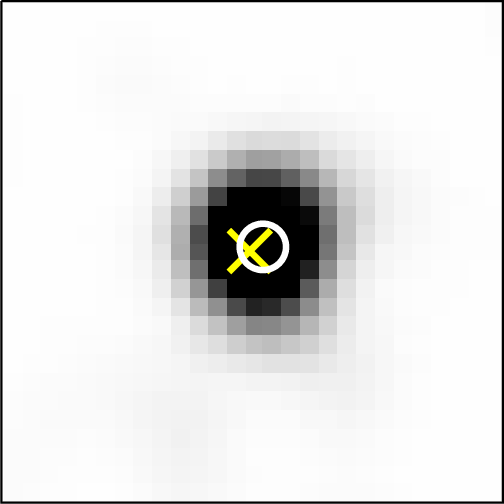}}
    \includegraphics[width=0.32\textwidth, trim = {0 0.5cm 0 0.3cm},clip]{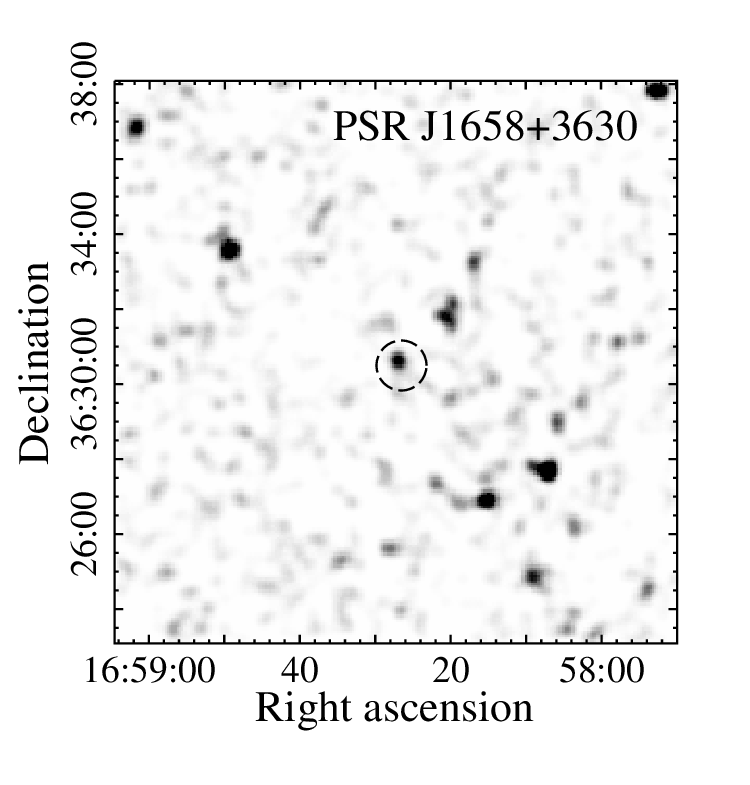}
    \put(-68,26) {\includegraphics[width=0.1\textwidth, clip=]{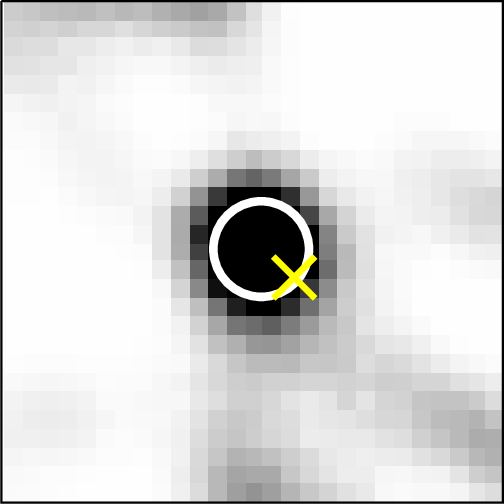}}
    \includegraphics[width=0.32\textwidth, trim = {0 0.5cm 0 0.3cm},clip]{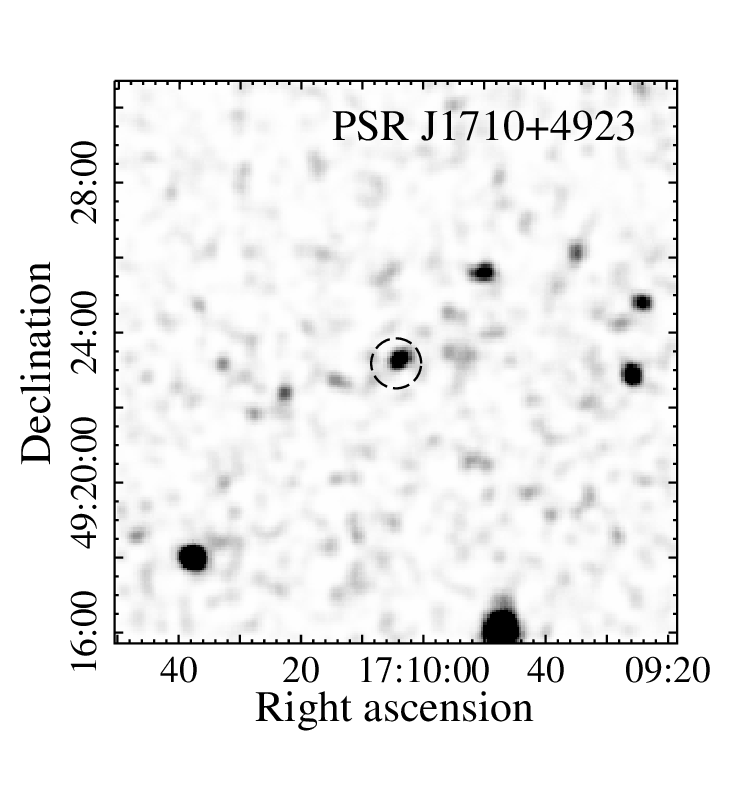}
    \put(-68,26) {\includegraphics[width=0.1\textwidth, clip=]{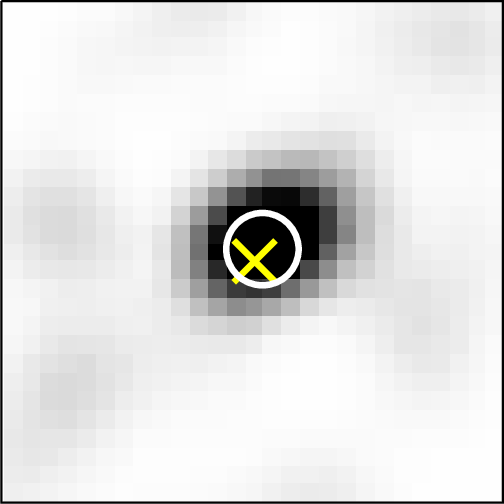}} \\
    \includegraphics[width=0.32\textwidth, trim = {0 0.5cm 0 0.3cm},clip]{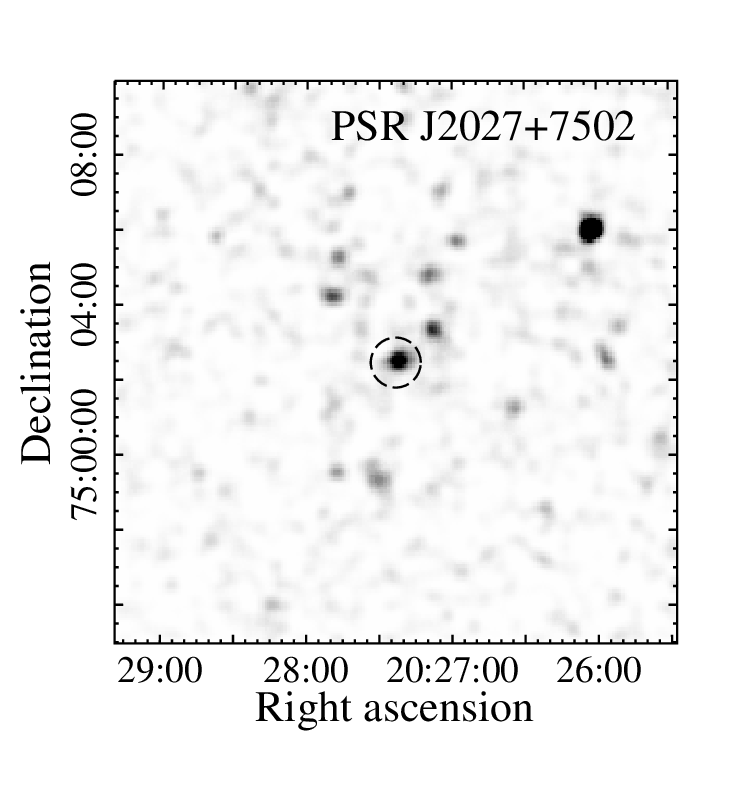}
    \put(-68,26) {\includegraphics[width=0.1\textwidth, clip=]{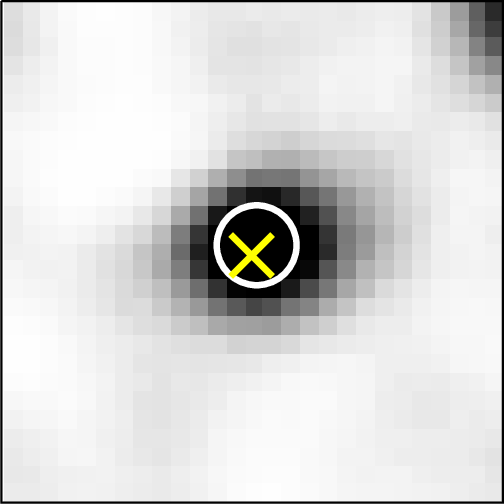}}
    \includegraphics[width=0.32\textwidth, trim = {0 0.5cm 0 0.3cm},clip]{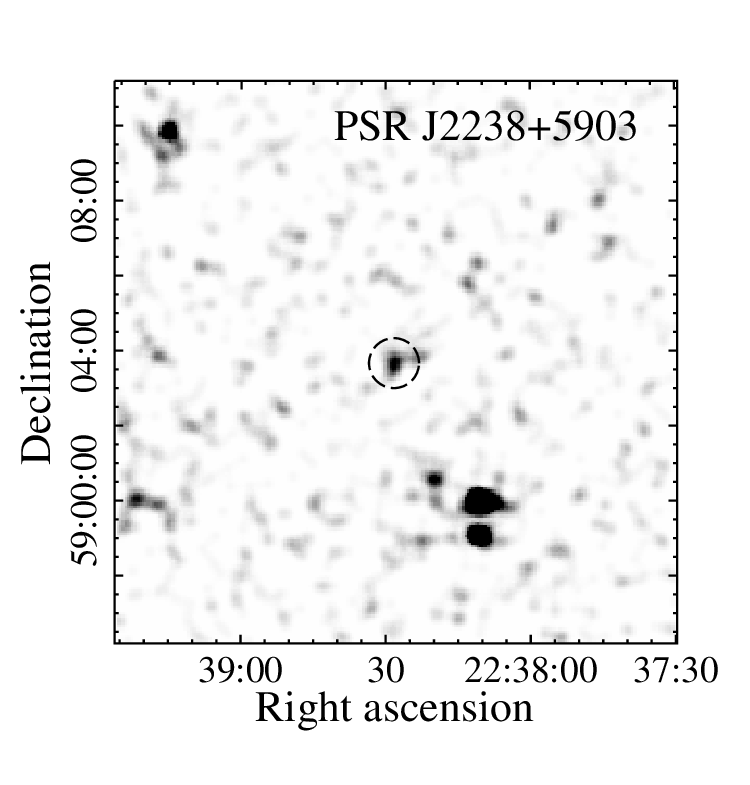}
    \put(-68,26) {\includegraphics[width=0.1\textwidth, clip=]{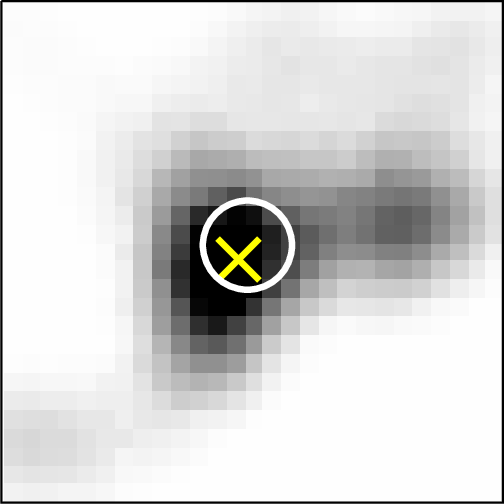}}
    \includegraphics[width=0.32\textwidth, trim = {0 0.5cm 0 0.3cm},clip]{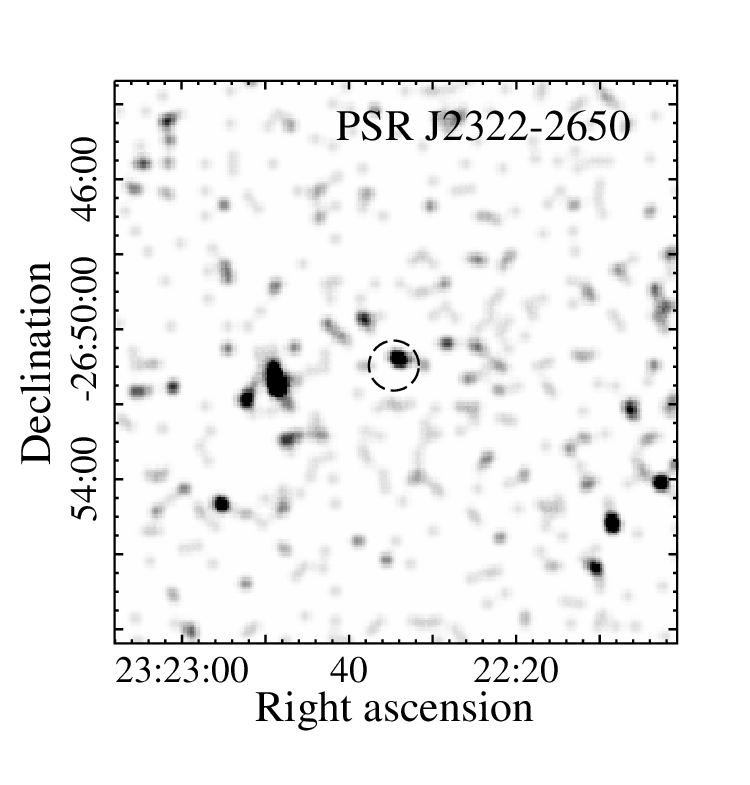}
    \put(-68,26) {\includegraphics[width=0.1\textwidth, clip=]{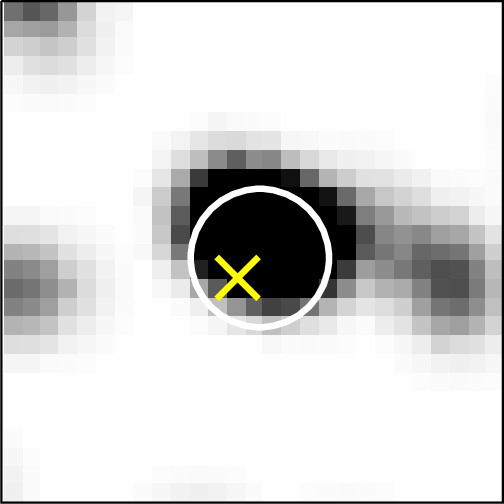}}
    \caption{15\amin\ $\times$ 15\amin\ images of the pulsars' fields as seen by eROSITA in the 0.3--2.3 keV range. 
    The pulsars' positions are shown by the dashed circles with the radius of 40\asec. 
    In the 1\farcm8~$\times$~1\farcm8 insets the white circles show the position uncertainties of the X-ray sources while the pulsars' positions are marked with the yellow crosses.}
    \label{fields}
\end{center}    
\end{figure*}
%---------------------------------------------------------------

%---------------------------------------------------------------
\begin{table*}
\caption{Parameters of the pulsars identified in X-rays taken from the ATNF catalogue.}
\begin{center}
\label{tab:pars}
\begin{tabular}{lcccccccc} \hline
N & PSR         & $P$, ms & $\dot{P}$, s s$^{-1}$  & $\tau_{\rm c}$, yr & $\dot{E}$, \ergs\ & $B$, G & DM, \dm & $D$, kpc \\ 
\hline 
1 & J0006$+$1834  & 693.7 & 2.1$\times$10$^{-15}$  & 5.24$\times$10$^6$   & $0.25\times10^{33}$  & 1.2$\times$10$^{12}$ & 11.4  & 0.86$^{\rm r}$ \\
2 & J0242$+$6256  & 591.7 & 8.0$\times$10$^{-15}$  & 1.16$\times$10$^6$   & $1.5\times10^{33}$   & 2.2$\times$10$^{12}$ & 3.9   & 0.21$^{\rm r}$  \\ 
3 & J0248$+$6021  & 217.1 &  5.5$\times$10$^{-14}$ & 6.24$\times$10$^4$   & $210\times10^{33}$   & 3.5$\times$10$^{12}$ &  370  & 2$^*$  \\ 
4 & J1125$+$7819  & 4.2   & 6.94$\times$10$^{-21}$ & 9.59$\times$10$^9$   & $3.7\times10^{33}$  & 1.73 $\times$10$^8$  & 12.03 & 0.90$^{\rm r}$ \\
5 & J1544$+$4937  & 2.1   & 2.79$\times$10$^{-21}$ & 12.2$\times$10$^9$   & $11\times10^{33}$   & 7.86$\times$10$^7$   & 23.2  & 2.98$^{\rm r}$ \\
6 & J1602$+$3901  & 3.7   & --                     & --                   & --    & --                   & 17.26 & 1.63$^{\rm r}$ \\
7 & J1630$+$3734  & 3.3   & 1.08$\times$10$^{-20}$ & 4.9$\times$10$^9$    & $12\times10^{33}$    & 2$\times$10$^8$      & 14.2  & 1.19$^{\rm r}$ \\
8 & J1658$+$3630  & 33.0  & 11.6$\times$10$^{-20}$ & 4.5$\times$10$^9$    & $0.13\times10^{33}$  & 20$\times$10$^8$     & 3.04  & 0.22$^{\rm r}$ \\
9 & J1710$+$4923  & 3.2   & 1.82$\times$10$^{-20}$ & 2.8$\times$10$^9$    & $21.5\times10^{33}$  & 2.4$\times$10$^8$    & 7.08  & 0.50$^{\rm r}$ \\
10 & J2027$+$7502 & 515.2 & 0.89$\times$10$^{-15}$ & 9.17$\times$10$^6$   & $0.26\times10^{33}$  & 0.7$\times$10$^{12}$ & 11.4  & 0.74$^{\rm r}$ \\
11 & J2238$+$5903 & 162.7 & 9.7$\times$10$^{-14}$  & 2.7$\times$10$^4$    & $890\times10^{33}$   & 4.0$\times$10$^{12}$ & --    & 1.9$^{\gamma}$ \\
12 & J2322$-$2650 & 3.46  & 5.82$\times$10$^{-22}$ & 9.4$\times$10$^{10}$ & $0.55\times10^{33}$  & 4.55$\times$10$^7$   & 6.15  & 0.23$^*$ \\
\hline
\end{tabular}
\end{center}
\begin{tablenotes}
\item \footnotesize{\textbf{Notes.} $P$ is the pulsar spin period, $\dot{P}$ is the period derivative, $\tau_{\rm c}$ is the characteristic age, $\dot{E}$ is the spin-down luminosity, $B$ is the magnetic field at the equator, $D$ is the distance.
\item $^{\rm r}$ The DM distance determined using the YMW16 model. 
\item $^{\gamma}$ The pseudo-distance for radio-quiet $\gamma$-ray pulsars.
\item $^*$ See text.}
\end{tablenotes}

\end{table*}
%---------------------------------------------------------------

%For pulsar identification, we used the eROSITA catalog of X‑ray sources. \daz{
We used the eROSITA catalog of X-ray sources for pulsar identification.%} 
The latter was produced by the research team of the Russian eROSITA consortium for detection, identification, and compilation of X‑ray sources (headed by M.R. Gilfanov). 
We used the catalog which is based on the combined data from all sky surveys and includes sources with a detection likelihood $L$ of at least 6 (which corresponds to a confidence level of $\sim 3\sigma$ for the Gaussian distribution).

%%%%%%%%%%%%%%%%%%%%%%%%%%%%%%%%%%%%%%%%%%%%%%%%%%%%%%%%%%%%%%%%%%%%
%%%%%%%%%%%%%%%%%%%%%%%%%%%%%%%%%%%%%%%%%%%%%%%%%%%%%%%%%%%%%%%%%%%%

\subsection{Search results}

As a result of correlation of the pulsar list with the eROSITA source catalog, we discovered new candidate X-ray counterparts to 12 pulsars from our sample. 
These include two isolated middle-aged ($<1$ Myr) pulsars, three isolated old ($>1$ Myr) pulsars, including one rotating radio transient (RRAT), and six millisecond pulsars (MSPs), only one of which is isolated. 
The parameters of all 12 pulsars are presented in Table~\ref{tab:pars}, while the coordinates of the pulsars and their likely counterparts from the eROSITA catalog are given in Table~\ref{tab:x-ray}. 
The parameters of the eROSITA X‑ray sources are provided in Table~\ref{tab:x-ray-obs}. 
For convenience, the numbering of the pulsars is the same in all three tables. 
Note that the X‑ray fluxes in Table~\ref{tab:x-ray-obs} are taken from the eROSITA catalog. 
They are calculated for a power-law (PL) spectrum with a photon index of 2 and are not corrected for the interstellar absorption. 
These values may differ from those presented below for the objects the number of counts from which was sufficient for spectral analysis.
The eROSITA images of the pulsars' fields are presented in Fig.~\ref{fields}.

The distances to the radio pulsars listed in Table~\ref{tab:pars} (except for PSR J0248$+$6021 and PSR J2322$-$2650, see below) were determined from the DM using the YMW16 model \citep{ymw} for the distribution of free electrons in the Galaxy. 
For the radio-quiet $\gamma$-ray pulsar PSR J2238$+$5903, the so-called pseudo-distance was calculated. 
It is derived using the empirical relation between the spin-down luminosity $\dot{E}$ and the $\gamma$-ray flux $G_{100}$ \citep{sazparkinson}:

\begin{equation}
D^\gamma = 1.6\times\dot{E}_{34}^{0.25}G_{100,-11}^{-0.5}~\text{kpc},
\end{equation}
where $\dot{E_{34}}\equiv \dot{E}/(10^{34}$ \ergs), $G_{100,-11}\equiv G_{100}/(10^{-11}$ \flux).

%%%%%%%%%%%%%%%%%%%%%%%%%%%%%%%%%%%%%%%%%%%%%%%%%%%%%%%%%%%%%%%%%%%%
%%%%%%%%%%%%%%%%%%%%%%%%%%%%%%%%%%%%%%%%%%%%%%%%%%%%%%%%%%%%%%%%%%%%

\subsection{Data from other X-ray telescopes and spectral analysis}

We сhecked whether the found counterparts are presented in the Master X-Ray Catalog\footnote{\url{https://heasarc.gsfc.nasa.gov/w3browse/all/xray.html}} (XRAY) and in the Swift telescope data. 
As a result, two sources were detected by Swift, two -- by Chandra, and one -- by ROSAT. 

For the eROSITA sources with a detection likelihood $L>10$, we performed more detailed spectral analysis.
From the eROSITA data, spectra of the sources were extracted using a 60\asec-radius aperture while for the background we used an annulus with inner and outer radii of 120\asec and 300\asec. 
Any background objects falling into this region were excluded with 40\asec-radius circles. 
For pulsars detected by Swift, spectra were obtained using the online Swift/XRT data product generator\footnote{\url{https://www.swift.ac.uk/user\_objects/}} \citep{evans2009}.
The Chandra data were processed using the CIAO v.4.16 package. 
We applied the \texttt{chandra\_repro} tool to reprocess the data and the \texttt{specextract} routine to extract the spectra.

The eROSITA spectra were fitted in the 0.3--9 keV range, while the Chandra and Swift data were fitted in the 0.3--10 keV range using the XSPEC v.12.13.1 package\footnote{X-Ray Spectral Fitting Package \citep{xspec}}. 
In the case of data from several telescopes, we fitted the spectra simultaneously. 
To describe the non-thermal emission from the NS magnetosphere, a PL model was applied. 
This model was used as the primary fit for all spectra. 
If the resulting photon index was $>$3.5, indicating the possible presence of a thermal spectral component from the NS surface, we utilised either the black body (BB) model or the NSMAXG model, describing the emission from a magnetized hydrogen NS atmosphere \citep{ho2014}. 
For the latter, the NS mass and radius were fixed at 1.4~M$_\odot$ and 13 km, respectively, corresponding to the gravitational redshift of $1 + z_g = 1.21$.

Since the number of counts from the detected sources is low (see Table~\ref{tab:x-ray-obs}), spectra were grouped to ensure at least one count per energy bin, and the $C$-statistics \citep{cash} was chosen. 
The uncertainties of the fitted parameters presented below correspond to $1\sigma$ confidence intervals.

The TBABS model with WILM abundances \citep{wilm} was used to account for the interstellar absorption. 
For each pulsar, we estimated the column density $N_{\rm H}$ using three-dimensional maps of the Galactic interstellar extinction from Green et al.\footnote{\url{http://argonaut.skymaps.info/} \citep{green}}, or Lallement et al.\footnote{\url{https://stilism.obspm.fr/} \citep{lallement2014,lallement2018,capitanio}}, or the online calculator\footnote{\url{https://cxc.harvard.edu/toolkit/colden.jsp}}, and the distances from Table~\ref{tab:pars}. 
The reddening values obtained from the maps were converted to $N_{\rm H}$ utilizing the empirical relation from \citet{foight}. 
The resulting $N_{\rm H}$ values were fixed during spectral fitting. 
The unabsorbed fluxes $F_X$ provided below are given for the 0.3--10 keV range unless stated otherwise.

%---------------------------------------------------------------
\begin{table*}
\caption{Candidate counterparts to the pulsars from the eROSITA catalogue.}
\label{tab:x-ray}
\begin{center}
\begin{tabular}{lccccc} \hline
N & X-ray source      & $\alpha_{\rm psr}$ & $\delta_{\rm psr}$    & $\Delta$, arcsec & $R_{98}$, arcsec \\ 
\hline
1 & SRGe J000604.8$+$183451   & 00:06:04.8(2)      & $+$18:34:59(4)        & 8.5              & 9.9 \\
2 & SRGe J024240.5$+$625642   & 02:42:39.61        & $+$62:56:42.44        & 5.9              & 9.3 \\ 
3 & SRGe J024817.7$+$602132   & 02:48:18.617(1)    & $+$60:21:34.72(1)     & 7.4              & 10.9\\ 
4 & SRGe J112600.5$+$781948   & 11:25:59.85664(7)  & $+$78:19:48.71546(19) & 2.2              & 14.5 \\ 
5 & SRGe J154404.7$+$493759   & 15:44:04.4911(1)   & $+$49:37:55.374(1)    & 4.2              & 12.1 \\  
6 & SRGe J160218.8$+$390106   & 16:02:18.84(8)     & $+$39:01:01.8(9)      & 4.3              & 9.4 \\
7 & SRGe J163036.2$+$373443   & 16:30:36.46693(7)  & $+$37:34:42.097(1)    & 3.0              & 5.0 \\
8 & SRGe J165827.1$+$363036   & 16:58:26.5198(3)   & $+$36:30:30.095(3)    & 9.4              & 10.3 \\
9 & SRGe J171004.3$+$492314   & 17:10:04.44109(2)  & $+$49:23:11.4784(3)   & 3.1              & 7.8 \\
10 & SRGe J202723.0$+$750232  & 20:27:23.27(2)     & $+$75:02:29.3(2)      & 2.5              & 8.6 \\
11 & SRGe J223828.0$+$590344  & 22:38:28.27(4)     & $+$59:03:40.8(4)      & 3.7              & 9.6 \\
12 & SRGe J232234.3$-$265054  & 23:22:34.638818(5) & $-$26:50:58.38497(10) & 6.5              & 14.9  \\ 
\hline
\end{tabular}
\end{center}
\begin{tablenotes}
\item \footnotesize{\textbf{Notes.} $\alpha_{\rm psr}$ and $\delta_{\rm psr}$ are the right ascension and declination of the pulsar from the radio or $\gamma$-ray data. 
They are taken from the ATNF catalogue, except for coordinates of PSR J0242$+$6256, for which the updated data were kindly provided by J. Hessels (private communication).
Numbers in parentheses denote $1\sigma$ coordinate uncertainties relating to the last significant digit quoted.
$\Delta$ is the offset between the pulsar and the X-ray source positions.
$R_{98}$ is the 98\% positional uncertainty in X-rays. }
\end{tablenotes}
\end{table*}
%---------------------------------------------------------------

%%%%%%%%%%%%%%%%%%%%%%%%%%%%%%%%%%%%%%%%%%%%%%%%%%%%%%%%%%%%%%%%%%%%
%%%%%%%%%%%%%%%%%%%%%%%%%%%%%%%%%%%%%%%%%%%%%%%%%%%%%%%%%%%%%%%%%%%%
%%%%%%%%%%%%%%%%%%%%%%%%%%%%%%%%%%%%%%%%%%%%%%%%%%%%%%%%%%%%%%%%%%%%
%%%%%%%%%%%%%%%%%%%%%%%%%%%%%%%%%%%%%%%%%%%%%%%%%%%%%%%%%%%%%%%%%%%%

\section{Properties of the pulsars and their likely counterparts}

Let us consider the pulsars in the order in which they are presented in Tables~\ref{tab:pars}--\ref{tab:x-ray-obs}.

%%%%%%%%%%%%%%%%%%%%%%%%%%%%%%%%%%%%%%%%%%%%%%%%%%%%%%%%%%%%%%%%%%%%
%%%%%%%%%%%%%%%%%%%%%%%%%%%%%%%%%%%%%%%%%%%%%%%%%%%%%%%%%%%%%%%%%%%%

\subsection{1. PSR J0006$+$1834}

PSR J0006$+$1834 was discovered by the Arecibo radio telescope \citep{Camilo&Nice}. 
The DM distance to the pulsar is $\approx 0.9$ kpc. 
In this case, the pulsar should be 600 pc away from the Galactic plane. 
However, the majority of pulsars is located in the disk, so this distance estimate may be considered as an upper limit. 
In any case, according to the map by Green et al., the total interstellar absorption in the pulsar direction is low,  $N_{\rm H} = 3.5\times10^{20}$~см$^{-2}$.
We fitted the J0006+1834 spectrum with the PL model. 
Due to large uncertainties of the fit parameters, we fixed the photon index at $\Gamma = 2$, which is the typical value for pulsars \citep{kargaltsev&pavlov}. 
The resulting unabsorbed flux is $F_X = 1.3^{+1.3}_{-1.1} \times 10^{-14}$ \flux\ and $C = 19$ for 18 degrees of freedom (d.o.f.).

%%%%%%%%%%%%%%%%%%%%%%%%%%%%%%%%%%%%%%%%%%%%%%%%%%%%%%%%%%%%%%%%%%%%
%%%%%%%%%%%%%%%%%%%%%%%%%%%%%%%%%%%%%%%%%%%%%%%%%%%%%%%%%%%%%%%%%%%%

\subsection{2. PSR J0242$+$6256}

PSR J0242$+$6256 was discovered by the Green Bank Telescope in the GBT350 radio survey \citep{hessels2008}. 
It emits sporadic radio bursts and most likely belongs to the RRAT class.

The likely X-ray counterpart to the pulsar was detected by eROSITA and Swift.
The column density in the pulsar direction is highly uncertain.
The low DM $\approx 4$ \dm\ corresponds to $N_{\rm H}=1.2\times 10^{20}$ cm$^{-2}$ \citep{he2013}.
However, according to the extinction map by Lallement et al., $N_{\rm H}=7.7\times 10^{20}$ cm$^{-2}$.
This discrepancy can be explained by a low ionization ratio along the line of sight or by inaccuracies in the extinction map.

We fitted the spectra obtained from the Swift and eROSITA data simultaneously.
For $N_{\rm H} = 1.2\times 10^{20}$ cm$^{-2}$, the photon index is $\Gamma=1.4^{+0.5}_{-0.4}$, the unabsorbed flux is $F_X=7^{+6}_{-3}\times 10^{-14}$ \flux, and $C$/d.o.f. = 33/50.
For $N_{\rm H}=7.7\times 10^{20}$ cm$^{-2}$, $\Gamma=1.8\pm0.5$, $F_X=6^{+5}_{-2}\times 10^{-14}$ \flux, and $C$/d.o.f. = 33/50.

%%%%%%%%%%%%%%%%%%%%%%%%%%%%%%%%%%%%%%%%%%%%%%%%%%%%%%%%%%%%%%%%%%%%
%%%%%%%%%%%%%%%%%%%%%%%%%%%%%%%%%%%%%%%%%%%%%%%%%%%%%%%%%%%%%%%%%%%%

\subsection{3. PSR J0248$+$6021}

PSR J0248$+$6021 was discovered in the Nan{\c c}ay radio telescope survey \citep{foster1997}.
It was also detected in $\gamma$-rays \citep{abdo2010}. 
According to \citet{Theureau}, the pulsar is most likely located in the giant H~II region W5 at a distance of 2 kpc, which explains its very high DM. 
An association with the TeV source LHAASO J0248$+$6021 was also suggested \citep{psrj0248-lhaaso}.

\citet{psrj0248-Mignani} reported that the pulsar was not detected in Chandra data (obsID 13289). 
However, the XRAY catalog and the second Chandra catalog (2CXO) contain a point source at the pulsar's position, 2CXO J024818.6$+$602134, with a flux of $3.9^{+2.1}_{-2.2}\times10^{-15}$ erg cm$^{-2}$ s$^{-1}$ in the 0.5--7 keV range.
The latter is consistent within uncertainties with the eROSITA data.

%%%%%%%%%%%%%%%%%%%%%%%%%%%%%%%%%%%%%%%%%%%%%%%%%%%%%%%%%%%%%%%%%%%%
%%%%%%%%%%%%%%%%%%%%%%%%%%%%%%%%%%%%%%%%%%%%%%%%%%%%%%%%%%%%%%%%%%%%

\subsection{4. PSR J1125$+$7819}

The MSP PSR J1125+7819 was discovered in the radio by the Green Bank Telescope \citep{stovall2014}. 
It is in a binary system with a He-core white dwarf \citep{lynch}. 
The orbital period is 15.36 d.

%---------------------------------------------------------------
\begin{table*}
\caption{Parameters of the candidate counterparts from the eROSITA catalogue.}
\label{tab:x-ray-obs}
\begin{center}
\begin{tabular}{lccccc} \hline
N  & PSR                & $T_{\rm exp}$, s & $N_{\rm cts}$ & $f_X$, \flux               & $L$ \\
\hline
1  & J0006$+$1834       & 682              & 10.6$\pm$3.8  & 1.42(50)$\times$10$^{-14}$ & 13  \\
2  & J0242$+$6256$^{*}$ & 1038             & 21.3$\pm$5.7  & 1.87(50)$\times$10$^{-14}$ & 21 \\
3  & J0248$+$6021$^{*}$ & 1019             & 11.5$\pm$4.4  & 1.03(39)$\times$10$^{-14}$ & 7 \\
4  & J1125$+$7819       & 792              & 9.9$\pm$4.1   & 1.15(48)$\times$10$^{-14}$ & 7\\
5  & J1544$+$4937       & 1759             & 11.1$\pm$4.5  & 0.57(23)$\times$10$^{-14}$ & 6 \\
6  & J1602$+$3901       & 1383             & 19.9$\pm$5.7  & 1.31(38)$\times$10$^{-14}$ & 18 \\
7  & J1630$+$3734$^{*}$ & 1305             & 68.5$\pm$9.3  & 4.79(65)$\times$10$^{-14}$ & 116\\
8  & J1658$+$3630       & 1193             & 12.9$\pm$4.5  & 1.00(34)$\times$10$^{-14}$ & 11 \\
9  & J1710$+$4923       & 2372             & 42.5$\pm$8.1  & 1.64(31)$\times$10$^{-14}$ & 42 \\
10 & J2027$+$7502       & 2883             & 41.6$\pm$8.1  & 1.32(26)$\times$10$^{-14}$ & 39 \\
11 & J2238$+$5903$^{*}$ & 1260             & 16.3$\pm$5.2  & 1.18(37)$\times$10$^{-14}$ & 13 \\
12 & J2322$-$2650       & 348              & 7.3$\pm$3.05  & 1.92(80)$\times$10$^{-14}$ & 8  \\
\hline
\end{tabular}
\end{center}
\begin{tablenotes}
\item \footnotesize{\textbf{Notes.} $T_{\rm exp}$ is the vignetting-corrected exposure time. 
$N_{\rm cts}$ is the background-corrected number of counts in the 0.3--2.3 keV range. 
The fluxes $f_X$ were calculated in the 0.3--2.3 keV range assuming the power-law spectra with the photon index $\Gamma = 2$ and column density $N_{\rm H} = 3\times 10^{20}$ cm$^{-2}$ and were not corrected for the interstellar absorption.  
$L$ is the detection likelihood of the X-ray source, $L=-\text{ln}p$, where $p$ is the probability of the source being a background fluctuation. \\
$^{*}$ The source was also detected in other X-ray catalogues. }
\end{tablenotes}
\end{table*}
%---------------------------------------------------------------

%%%%%%%%%%%%%%%%%%%%%%%%%%%%%%%%%%%%%%%%%%%%%%%%%%%%%%%%%%%%%%%%%%%%
%%%%%%%%%%%%%%%%%%%%%%%%%%%%%%%%%%%%%%%%%%%%%%%%%%%%%%%%%%%%%%%%%%%%

\subsection{5. PSR J1544$+$4937}

PSR J1544$+$4937 was first identified by the Giant Metrewave Radio Telescope during pulsar searches among unidentified $\gamma$-ray sources \citep{1544gmrt}. 
Pulsations with the pulsar's spin period were also detected in $\gamma$-rays. 
J1544$+$4937 is an eclipsing radio pulsar in a ``black widow'' system, a subclass of ``spider'' pulsars, which represent close binary systems where a low-mass stellar companion is irradiated and ablated by the pulsar wind. 
The orbital period of J1544$+$4937 is 2.9 h. 
\citet{tang2014} discovered its optical counterpart. 
Modeling the object's optical light curves obtained with the Gran Telescopio Canarias, \citet{matasanchez} derived the distance to the pulsar of $3.1^{+0.5}_{-0.4}$ kpc, which is close to the estimate from the DM (see Table ~\ref{tab:pars}).

%%%%%%%%%%%%%%%%%%%%%%%%%%%%%%%%%%%%%%%%%%%%%%%%%%%%%%%%%%%%%%%%%%%%
%%%%%%%%%%%%%%%%%%%%%%%%%%%%%%%%%%%%%%%%%%%%%%%%%%%%%%%%%%%%%%%%%%%%

\subsection{6. PSR J1602$+$3901}

The MSP PSR J1602$+$3901 was recently discovered in the radio in the Low-Frequency Array (LOFAR) Two-metre Sky Survey \citep[LoTSS][]{sobey2022}.
It forms a binary system with a white dwarf and an orbital period of 6.32 d.

According to the map by Green et al., the interstellar absorption in the pulsar direction is low, $N_{\rm H}=6.1\times10^{20}$~cm$^{-2}$. 
Fitting the spectrum with the PL model, we obtained the photon index $\Gamma = 3.2 \pm 0.6$, unabsorbed flux $F_X = (3.4 \pm 0.9) \times 10^{-14}$ \flux\ and $C$/d.o.f. = 31/56. 

%%%%%%%%%%%%%%%%%%%%%%%%%%%%%%%%%%%%%%%%%%%%%%%%%%%%%%%%%%%%%%%%%%%%
%%%%%%%%%%%%%%%%%%%%%%%%%%%%%%%%%%%%%%%%%%%%%%%%%%%%%%%%%%%%%%%%%%%%

\subsection{7. PSR J1630$+$3734}

PSR J1630$+$3734 is a binary MSP. 
It was discovered in the radio observations of unidentified $\gamma$-ray sources, and pulsations were also detected in $\gamma$-rays \citep{ray2012,sanpaarsa}. 
The pulsar's companion is a white dwarf, which was detected in the optical with the Gran Telescopio Canarias \citep{kirichenko2020}. The orbital period of the system is 12.53 d.

According to the XRAY catalog, the pulsar was also detected by ROSAT, with the count rate of $4.4 \pm 1.1$ cts~ks$^{-1}$ and the exposure time of 6.86 ks. 
In addition, the pulsar's field was observed 4.9 ks by Swift, but it was not detected.

We extracted the source spectrum from the eROSITA data and fitted it with a PL model.
The absorption in the pulsar direction is very low\footnote{According to Colden: Galactic Neutral Hydrogen Density Calculator}, $N_{\rm H} \approx 10^{20}$ см$^{-2}$.
We obtained the photon index $\Gamma=2.3 \pm 0.3$, unabsorbed flux $F_X=7.7^{+2.1}_{-1.3}\times10^{-14}$ \flux\ and $C$/d.o.f. = 97/83.

%%%%%%%%%%%%%%%%%%%%%%%%%%%%%%%%%%%%%%%%%%%%%%%%%%%%%%%%%%%%%%%%%%%%
%%%%%%%%%%%%%%%%%%%%%%%%%%%%%%%%%%%%%%%%%%%%%%%%%%%%%%%%%%%%%%%%%%%%

\subsection{8. PSR J1658$+$3630}

The binary PSR J1658$+$3630 was discovered by the LOFAR radio telescope and belongs to the mildly recycled pulsar population \citep{lofar2020}. 
Its companion, likely a white dwarf with a CO or ONeMg core, was identified in the SDSS catalog. 
The orbital period of the system is 3.016 d.

The maximum interstellar absorption in the pulsar direction, according to the map by Green et al., is $N_{\rm H} = 3.5\times10^{20}$ cm$^{-2}$. 
Due to the large uncertainties of the fit parameters, we fixed the photon index at $\Gamma = 2$. 
As a result, we obtained the unabsorbed flux of $F_X=2.4^{+1.1}_{-1.0}\times10^{-14}$ \flux\ and $C$/d.o.f. = 47/46.

%%%%%%%%%%%%%%%%%%%%%%%%%%%%%%%%%%%%%%%%%%%%%%%%%%%%%%%%%%%%%%%%%%%%
%%%%%%%%%%%%%%%%%%%%%%%%%%%%%%%%%%%%%%%%%%%%%%%%%%%%%%%%%%%%%%%%%%%%

\subsection{9. PSR J1710$+$4923}

PSR J1710$+$4923 is an isolated MSP discovered in the radio by the Green Bank telescope \citep{stovall2014}.
Its field was observed for 13 ks with Swift, but the pulsar was not detected.

According to the map of Green et al., the maximal absorption in the pulsar direction is $\approx$~3.5~$\times$~10$^{20}$~cm$^{-2}$. 
The X-ray spectrum of the likely counterpart can be described by the PL model with the photon index $\Gamma=2.7 \pm 0.4$, unabsorbed flux $F_X=3.3^{+0.8}_{-0.6}\times10^{-14}$~\flux\ and $C/{\rm d.o.f.}=79/93$.

%%%%%%%%%%%%%%%%%%%%%%%%%%%%%%%%%%%%%%%%%%%%%%%%%%%%%%%%%%%%%%%%%%%%
%%%%%%%%%%%%%%%%%%%%%%%%%%%%%%%%%%%%%%%%%%%%%%%%%%%%%%%%%%%%%%%%%%%%

\subsection{10. PSR J2027$+$7502}

PSR J2027$+$7502 was recently discovered in the radio survey of the Green Bank telescope \citep{lynch}. 

The column density at the distance of 0.8 kpc in the pulsar direction, according to the map of Green et al., is $N_{\rm H} = 1.7 \times 10^{21}$ cm$^{-2}$.
The PL fit of the source spectrum resulted in $\Gamma=1.8\pm0.5$, $F_X=6^{+3}_{-2}\times10^{-14}$ \flux\ and $C/{\rm d.o.f.}=99/91$.

%%%%%%%%%%%%%%%%%%%%%%%%%%%%%%%%%%%%%%%%%%%%%%%%%%%%%%%%%%%%%%%%%%%%
%%%%%%%%%%%%%%%%%%%%%%%%%%%%%%%%%%%%%%%%%%%%%%%%%%%%%%%%%%%%%%%%%%%%

\subsection*{11. PSR J2238$+$5903}

PSR J2238$+$5903 is a radio-quiet $\gamma$-ray pulsar discovered by the Fermi telescope \citep{abdo2009}.
Its coordinates presented in Table~\ref{tab:x-ray} were obtained from the $\gamma$-ray data with the positional uncertainty of 3\asec\ \citep{Ray2011}.
The pulsar is possibly associated with the TeV source 1LHAASO J2238$+$5900 \citep{cao2024}.

J2238$+$5903 was observed in X-rays for about 10 ks with Chandra.
However, according to \citet{prinz&becker}, it was not detected.
Nevertheless, eROSITA found a point-like source at the pulsar's position (Fig.~\ref{fields}).
We checked the Chandra data and also detected 27$\pm$6 cts from the source which is designated as X223827.95$+$590343.1 in the XRAY catalogue and coincides with the J2238$+$5903 position. 
In addition, the pulsar's field was observed with Swift but it was not detected.

The column density for the distance of 1.9 kpc, $N_{\rm H} = 6.5 \times 10^{21}$ см$^{-2}$, was obtained from the map by Green et al.
We fitted the eROSITA and Chandra spectra simultaneously  with the PL model and found that the source is very soft: the photon index $\Gamma=4.0\pm0.4$.  
The BB model resulted in the effective temperature $T=216^{+29}_{-23}$ eV, emitting area radius $R=0.35^{+0.15}_{-0.11}D_{\rm 1.9\ kpc}$ km ($D_{\rm 1.9\ kpc} \equiv D/1.9\ \text{kpc}$), unabsorbed flux $F_X=7.4^{+2.1}_{-1.7}\times10^{-14}$ \flux\ and $C/{\rm d.o.f.}=69/70$.
Thus, emission originates from the hot spot on the NS surface.

We also tried the NSMAXG model, assuming the magnetic filed of 4$\times$10$^{12}$ G and distance of 1.9 kpc. The resulting NS surface temperature, as seen by a distant observer, $T^\infty=56\pm1$ eV, flux $F_X=3.4^{+0.4}_{-0.3}\times10^{-14}$ \flux\ and $C/{\rm d.o.f.}=79/71$.

%%%%%%%%%%%%%%%%%%%%%%%%%%%%%%%%%%%%%%%%%%%%%%%%%%%%%%%%%%%%%%%%%%%%
%%%%%%%%%%%%%%%%%%%%%%%%%%%%%%%%%%%%%%%%%%%%%%%%%%%%%%%%%%%%%%%%%%%%

\subsection{12. PSR J2322$-$2650}

PSR J2322$-$2650 is a binary MSP discovered in the radio during the High Time Resolution Universe survey \citep{psrj2322}.
It has a companion with a mass of $7.6\times10^{-4}$~M$_\odot$ in a circular 7.75-hour orbit. 
The radio timing parallax is $4.4 \pm 1.2$ mas corresponding to the distance of $230^{+90}_{-50}$ pc.

%%%%%%%%%%%%%%%%%%%%%%%%%%%%%%%%%%%%%%%%%%%%%%%%%%%%%%%%%%%%%%%%%%%%
%%%%%%%%%%%%%%%%%%%%%%%%%%%%%%%%%%%%%%%%%%%%%%%%%%%%%%%%%%%%%%%%%%%%
%%%%%%%%%%%%%%%%%%%%%%%%%%%%%%%%%%%%%%%%%%%%%%%%%%%%%%%%%%%%%%%%%%%%
%%%%%%%%%%%%%%%%%%%%%%%%%%%%%%%%%%%%%%%%%%%%%%%%%%%%%%%%%%%%%%%%%%%%

\section{Discussion and conclusions}

We detected 12 new candidate X‑ray counterparts to pulsars in the eastern Galactic hemisphere. 
A comparable number (15 with $L \geq 5$)
of possible counterparts was found in the western half of the sky \citep{erosita-grmn}. In total, this accounts for $\approx12$\% of the number of known pulsars emitting in the X‑rays \citep{xu2025}.

Table~\ref{Known} presents the results of correlation of the pulsar list with the eROSITA catalog for those objects which had been previously detected by other X-ray telescopes. 
eROSITA detected X‑ray emission from 38 pulsars, 8 MSPs (either isolated or in binary systems with a white dwarf companion) and 9 MSPs in spider binary systems.

Most of the new pulsars have a low DM, which is natural, as this implies small values of the interstellar absorption and distances.

For some objects, the number of detected counts allowed us to analyze their X‑ray spectra obtained from eROSITA data. 
For the spectral analysis, we also used archival Chandra and Swift data.

Among the interesting objects, the pulsar J0242+6256 can be noted, as it may become the second RRAT emitting in X-rays.
While over a hundred of RRATs are known\footnote{\url{http://astro.phys.wvu.edu/rratalog/}}, X‑ray emission has been detected only for one of them, PSR J1819$-$1458 \citep{reynolds2006}.
It has a purely thermal spectrum, which can be described by the BB model with the broad absorption line at $\sim1$~keV \citep{mclaughlin}.

J1658+3630 can be the third mildly recycled pulsar detected in X‑rays after PSR J2222$-$0137 \citep[the spin period is 32.8 ms;][]{prinz&becker} and PSR J1439$-$5501 \citep[the spin period is 28.6 ms;][]{erosita-grmn}.

For PSR J2238$+$5903, we estimated the surface temperature to be $\approx 56$ eV. 
A similar temperature was obtained for the Vela pulsar \citep{ofengeim&zyuzin} with a comparable age (10 kyr) which is one of the coldest NSs of close ages \citep{potekhin2020}. 
By this property, J2238$+$5903 can be an analog of Vela pulsar.

Deeper observations are necessary to confirm the counterparts, in particular by detecting pulsations with pulsars' spin periods and clarifying the parameters of their X-ray emission.

%%%%%%%%%%%%%%%%%%%%%%%%%%%%%%%%%%%%%%%%%%%%%%%%%%%%%%%%%%%%%%%%%%%%
%%%%%%%%%%%%%%%%%%%%%%%%%%%%%%%%%%%%%%%%%%%%%%%%%%%%%%%%%%%%%%%%%%%%
%%%%%%%%%%%%%%%%%%%%%%%%%%%%%%%%%%%%%%%%%%%%%%%%%%%%%%%%%%%%%%%%%%%%
%%%%%%%%%%%%%%%%%%%%%%%%%%%%%%%%%%%%%%%%%%%%%%%%%%%%%%%%%%%%%%%%%%%%

\subsection{Acknowledgments}

\footnotesize{This work used data obtained with eROSITA telescope onboard SRG observatory. The SRG observatory was built by Roskosmos in the interests of the Russian Academy of Sciences represented by its Space Research Institute (IKI) in the framework of the Russian Federal Space Program, with the participation of the Deutsches Zentrum f\"ur Luft- und Raumfahrt (DLR). The SRG/eROSITA X-ray telescope was built by a consortium of German Institutes led by the Max-Planck-Institut f\"ur extraterrestrische Physik (MPE), and supported by DLR. The SRG spacecraft was designed, built, launched and is operated by the Lavochkin Association and its subcontractors. The science data are downlinked via the Deep Space Network Antennae in Bear Lakes, Ussurijsk, and Baykonur, funded by Roskosmos. The eROSITA data used in this work were processed using the eSASS software system developed by the German eROSITA consortium and proprietary data reduction and analysis software developed by the Russian eROSITA Consortium. 
The work of YAS, AVK and DAZ (preparation of the pulsar sample for X-ray identification, search of counterparts in the XRAY catalog and Swift data, extraction of the spectra from the Chandra and Swift data, and analysis of the spectra of J0006$+$1834, J0242$+$6256, J1658$+$3630, J2027$+$7502, and J2238$+$5903) was supported by the baseline project FFUG-2024-0002 of the Ioffe Institute.
The analysis of the X-ray spectra of the millisecond pulsars J1602$+$3901, J1630$+$3734, and J1710$+$4923 by AVK was supported by the Russian Science Foundation project 22-12-00048-P.
DAZ thanks Pirinem School of Theoretical Physics for hospitality.}

%%%%%%%%%%%%%%%%%%%%%%%%%%%%%%%%%%%%%%%%%%%%%%%%%%%%%%%%%%%%
%%%%%%%%%%%%%%%%%%%%%%%%%%%%%%%%%%%%%%%%%%%%%%%%%%%%%%%%%%%%
%%%%%%%%%%%%%%%%%%%%%%%%%%%%%%%%%%%%%%%%%%%%%%%%%%%%%%%%%%%%
%%%%%%%%%%%%%%%%%%%%%%%%%%%%%%%%%%%%%%%%%%%%%%%%%%%%%%%%%%%%

\onecolumn

\normalsize{
\begin{center}
\begin{ThreePartTable}
\begin{TableNotes}
\item \footnotesize{\textbf{Notes.} The designations of the columns are the same as in Tables~\ref{tab:x-ray}--\ref{tab:x-ray-obs}.
\item $^{ext}$The source is extended due to the presence of the pulsar wind nebula.}
\end{TableNotes}
\begin{xltabular}{\textwidth}{lccccc}
\caption*{\footnotesize{\textbf{Table 4.} List of pulsars previously detected in X‑Rays.}} \label{Known} \\
\toprule
PSR                  & $\Delta$, arcsec  & $T_{\rm exp}$, s  & $N_{\rm cts}$      & $f_X$, \flux & $L$ \\
\midrule
\endfirsthead

\caption* {\footnotesize{\textbf{Table 4} -- continued.}}\\
\midrule
\endhead

\endfoot

\bottomrule
\insertTableNotes
\endlastfoot

J0002$+$6216           & 0.7        & 1296          &   33.96$\pm$6.96   &  2.33(48)$\times$10$^{-14}$ & 40  \\
J0007$+$7303$^{ext}$   & 8.7        & 1507          &  133.75$\pm$18.60  & 8(1)$\times$10$^{-14}$  & 46 \\  
B0114$+$58             & 2.8        & 1069          &   14.15$\pm$4.62   & 1.21(39)$\times$10$^{-14}$ &15 \\
J0205$+$6449           & 1.9        & 1069          &  432.57$\pm$34.64  & 37(3)$\times$10$^{-14}$   &126 \\
J0357$+$3205           & 3.1        & 811           &   31.09$\pm$6.27   & 5.51(71)$\times$10$^{-14}$   &51 \\
B0355$+$54             & 5.1        & 703           &   16.71$\pm$4.94   & 2.15(63)$\times$10$^{-14}$  &17\\
J0538$+$2817           & 2.5        & 528           &  420.94$\pm$21.51  &  72(4)$\times$10$^{-14}$  &1456\\
J0554$+$3107           & 2.4        & 481           &   18.29$\pm$5.19   & 3.47(98)$\times$10$^{-14}$ &21\\
J0622$+$3749           & 7.6        & 425           &   25.23$\pm$5.6    & 54(12)$\times$10$^{-15}$  &45    \\  
J1412$+$7922 (Calvera) & 1.2        & 1097          &  867.95$\pm$30.99  & 72(3)$\times$10$^{-14}$ &2886\\
RX J1605.3$+$3249      & 2.9        & 1144          & 3926.35$\pm$193.94 & 3.14(16)$\times$10$^{-12}$ &4159 \\
J1740$+$1000           & 3.1        & 640           &   80.87$\pm$10.38  & 1.16(15)$\times$10$^{-13}$  &118\\
J1741$-$2054$^{ext}$   & 2.0        & 527           &  195.87$\pm$14.99  & 3.39(26)$\times$10$^{-13}$  &486\\
J1809$-$1917           &  8.6       & 438           &   17.87$\pm$5.16   & 37(11)$\times$10$^{-15}$ &17 \\
J1809$-$2332           & 4.8        & 427           &   16.77$\pm$5.75   & 36(12)$\times$10$^{-15}$  &10\\
J1819$-$1458           & 0.5        & 423           &    87.0$\pm$10.51  & 1.89(23)$\times$10$^{-13}$ &159\\
B1822$-$09             & 9.5        & 426           &     9.9$\pm$3.99   & 2.12(86)$\times$10$^{-14}$&7\\
B1823$-$13             & 9.5        & 426           &    12.7$\pm$4.92   & 2.73(99)$\times$10$^{-14}$&8\\
J1833$-$1034$^{ext}$   & 1.2        & 416           & 1342.96$\pm$42.12  & 29.5(9)$\times$10$^{-13}$ &4121 \\
J1836$+$5925           & 3.0        & 2438          &   69.59$\pm$9.97   & 2.60(37)$\times$10$^{-14}$  &83 \\
J1838$-$0655           & 0.6        & 405           &   20.98$\pm$5.26   & 47(12)$\times$10$^{-15}$  &33 \\
J1846$-$0258           & 5.3        & 430           &   98.70$\pm$13.13  & 21(3)$\times$10$^{-14}$  &70\\
J1849$-$0001           & 3.8        & 443           &   17.76$\pm$4.71   & 3.68(98)$\times$10$^{-14}$ &29  \\
J1852$+$0040           & 3.7        & 445           &   62.65$\pm$12.21  & 13(2)$\times$10$^{-14}$ &25\\
J1930$+$1852$^{ext}$   & 2.9        & 476           &  282.77$\pm$20.33  & 54(4)$\times$10$^{-14}$  &529\\
B1929$+$10             & 4.8        & 431           &   61.58$\pm$8.63   & 12(2)$\times$10$^{-14}$ &133 \\
B1951$+$32$^{ext}$     & 0.9        & 584           &  483.59$\pm$31.56  & 76(5)$\times$10$^{-14}$  &229 \\
J1957$+$5033           & 1.0        & 942           &   55.50$\pm$10.21  &  5.39(99)$\times$10$^{-14}$ &35 \\
J1958$+$2846           & 7.8        & 532           &   11.47$\pm$4.35   & 1.97(75)$\times$10$^{-14}$&8 \\
J2016$+$3711$^{ext}$   & 10.3       & 942           &  105.78$\pm$16.97  & 17(3)$\times$10$^{-14}$  &41 \\
J2021$+$3651$^{ext}$   & 3.9        & 590           &   96.65$\pm$11.31  & 15(2)$\times$10$^{-14}$ &148\\
J2022$+$3842           & 3.7        & 611           &   13.16$\pm$4.96   & 1.97(74)$\times$10$^{-14}$ &7 \\
J2030$+$4415           & 12.2       & 700           &   12.05$\pm$4.42   & 1.57(58)$\times$10$^{-14}$ &9  \\
J2043$+$2740           & 10.7       & 473           &   11.60$\pm$4.16   & 2.22(79)$\times$10$^{-14}$ &12\\
RX J2143.0+0654        & 2.4        & 377           &  844.05$\pm$31.24  & 20.5(8)$\times$10$^{-13}$   &1983\\
B2224$+$65             & 5.6        & 1736          &   17.97$\pm$5.53   & 0.95(29)$\times$10$^{-14}$ &13\\
J2229$+$6114$^{ext}$   & 4.4        & 1453          &  215.66$\pm$18.02  & 14(1)$\times$10$^{-14}$ &271\\ 
B2334$+$61             & 5.3        & 1333          &   45.18$\pm$8.07   &  3.09(55)$\times$10$^{-14}$ &54\\
\midrule
\multicolumn{6}{c}{MSPs: isolated or those with a white dwarf companion} \\
\midrule 
J0030$+$0451          & 2.2         & 651           &  158.57$\pm$13.33  & 22(2)$\times$10$^{-14}$ &453 \\
J0218$+$4232          & 0.3         & 844           &   70.15$\pm$9.57   & 76(10)$\times$10$^{-15}$  &122 \\
J0337$+$1715          & 6.5         & 722           &     9.3$\pm$3.7    & 1.18(47)$\times$10$^{-14}$ &9 \\
J0740$+$6620          & 10.2        & 557           &     9.0$\pm$3.6    & 1.48(59)$\times$10$^{-14}$ &9 \\
J1012$+$5307          & 9.8         & 464           &   13.01$\pm$4.40   & 2.56(86)$\times$10$^{-14}$ &12\\
J1744$-$1134          & 8.0   & 538       &    10.4$\pm$4.26   & 1.77(72)$\times$10$^{-14}$&8 \\
J2124$-$3358          & 2.4         & 381           &   35.40$\pm$6.49   & 85(1)$\times$10$^{-14}$ &68 \\
J2302$+$4442          & 1.1         & 913           &   12.61$\pm$4.33   & 1.26(43)$\times$10$^{-14}$ &13\\
\midrule
\multicolumn{6}{c}{MSPs in spider systems} \\
\midrule 
J0212$+$5320          & 0.7         & 879           &  181.22$\pm$14.26  & 19(1)$\times$10$^{-14}$ &499 \\
J0636$+$5128          & 6.4         & 469           &   17.33$\pm$4.74   &  3.34(92)$\times$10$^{-14}$ &22\\
J1653$-$0158          & 0.5         & 575           &   26.00$\pm$6.23   & 4.13(99)$\times$10$^{-14}$ &26 \\
B1957$+$20            & 8.9         & 466           &   12.11$\pm$4.31   & 2.37(84)$\times$10$^{-14}$ &11\\
J2115$+$5448          & 3.9         & 1168          &   11.00$\pm$4.43   & 0.86(34)$\times$10$^{-14}$  &9 \\
J2129$-$0429          & 1.7         & 355           &   32.96$\pm$6.47   & 85(2)$\times$10$^{-14}$ &57 \\
J2214$+$3000          & 2.9         & 507           &   12.28$\pm$4.26   & 2.20(76)$\times$10$^{-14}$ &13\\
J2215$+$5135          & 5.9         & 1105          &   23.98$\pm$6.00   & 1.99(49)$\times$10$^{-14}$ &23\\
J2339$-$0533          & 1.2         & 432           &   24.78$\pm$5.59   & 52(12)$\times$10$^{-15}$  &40 \\
\end{xltabular}
\end{ThreePartTable}

\end{center}
}
\end{document}